\documentclass[aps,prl,final,twocolumn,superscriptaddress,floatfix]{revtex4-1}
\usepackage[latin9]{inputenc}
\usepackage{graphicx,epsf}
\usepackage{amssymb}
\usepackage{color}
\usepackage{amsmath}

\newcommand{\TKZ}{T_{\rm K}^{0}}
\newcommand{\TK}{T_{\rm K}}
\newcommand{\xik}{\xi_{\rm K}}
\newcommand{\xic}{\xi_{\rm c}}

\begin{document}


\title{
Quantum phase transition between one-channel and two-channel Kondo polarons 
}

\author{Juli\'an Rinc\'on}
\affiliation{Center for Nanophase Materials Sciences, Oak Ridge National Laboratory, Oak Ridge, TN 37831, USA}
\author{Daniel J. Garc\'{\i}a}
\author{K. Hallberg}
\affiliation{Instituto Balseiro, Centro At\'omico Bariloche, CNEA and CONICET, 8400 Bariloche, Argentina}
\author{Matthias Vojta}
\affiliation{Institut f\"ur Theoretische Physik, Technische Universit\"at Dresden, 01062 Dresden, Germany}

\date{\today}

\begin{abstract}
For a mobile spin-1/2 impurity, coupled antiferromagnetically to a one-dimensional gas of
fermions, perturbative ideas have been used to argue in favor of two-channel Kondo
behavior of the impurity spin. Here we combine general considerations and extensive
numerical simulations to show that the problem displays a novel quantum phase transition between
two-channel and one-channel Kondo screening upon increasing the Kondo coupling. We
construct a ground-state phase diagram and discuss the various non-trivial crossovers as
well as possible experimental realizations.
\end{abstract}


\maketitle


The problem of dilute particles moving in quantum liquids finds
realizations in diverse areas of modern physics~\cite{rosch}, such as charge carriers in
weakly doped semiconductors or Mott insulators, ions in $^3$He, muons in metals,
electrons in multi-band quantum wires, and multi-component ultracold gases with strong
population imbalance~\cite{palzer}.
For dilute particles with internal degree of freedom, e.g.\ spin, a connection to
quantum impurity problems, such as the Kondo effect, is natural. Indeed, a recent paper~\cite{lama_prl} argued that a spinful particle moving in a one-dimensional (1d) electron
gas creates a Kondo polaron which realizes the two-channel Kondo (2CK) effect.
This remarkable many-body effect occurs if a spin 1/2 is {\em overscreened} by
the coupling to two equivalent screening channels of conduction electrons, leading to
exotic non-Fermi liquid behavior~\cite{noz,andrei1984,aflu91}.
%
(For the 2CK polaron of Ref.~\onlinecite{lama_prl} the two screening channels are realized
by independent left-moving and right-moving fermions of the electron gas.)
While an unambiguous verification of 2CK behavior in solids containing magnetic ions is
still a challenge~\cite{coxzawa}, success was reported~\cite{potok} for a
nano-structured device consisting of a quantum dot with two reservoirs.

The 2CK effect is unstable w.r.t.\ channel asymmetry, such that the 2CK fixed point can
be understood as a critical point separating two single-channel Kondo (1CK) phases.
However, settings with a true quantum phase transition (QPT) \cite{ssbook,mvrev} between
1CK and 2CK phases are rare: the only example known to us is a proposal involving a
quantum dot coupled to helical edge states of a topological insulator~\cite{silotri}. In
contrast, for an impurity coupled to a standard Luttinger liquid, it has been argued
that either a 1CK or a 2CK phase is stable depending on the host's Luttinger
parameter~\cite{kondo_ll}, but a QPT upon varying an impurity parameter does not occur.

In this paper, we shall argue that the Kondo-polaron model of Ref.~\onlinecite{lama_prl}
realizes a novel QPT between 1CK and 2CK polarons.
We consider a single spin-1/2 particle, henceforth called ``impurity'', which moves
in a 1d gas of spin-1/2 fermions. The two species (or bands)
are coupled by an antiferromagnetic exchange interaction $J$.
The full lattice Hamiltonian reads:
\begin{eqnarray}
\mathcal{H} &=& \sum_{i=1}^L \sum_\sigma
\left(- t c_{i\sigma}^\dagger c_{i+1\sigma} - t' d_{i\sigma}^\dagger d_{i+1\sigma} + {\rm H.c.} \right)
\nonumber\\
& +& \sum_{i=1}^L
\left(J \vec{S}_{i} \cdot \vec{s}_{i} + h S_i^z \right),
\label{h}
\end{eqnarray}
where $n_i = \sum_\sigma c_{i\sigma}^\dagger c_{i\sigma}$ and  $\vec{s}_{i} =
\sum_{\sigma\sigma'} c_{i\sigma}^\dagger \vec{\tau}_{\sigma\sigma'} c_{i\sigma'}$, with
$\vec{\tau}$ the Pauli matrices, denote the local charge and spin densities,
respectively, of the conduction-band fermions. Their total density is given by
$n_c = \sum_i n_i/L$ and their bandwidth by $W=4t$. The impurity is described by $d$
operators, with local densities $N_i$ and $\vec{S}_{i}$ and the total filling fixed to
exactly one particle, $\sum_i N_i = 1$. We have also included a magnetic field $h$
coupling to the impurity.

\begin{figure}[!t]
\includegraphics[width=\columnwidth]{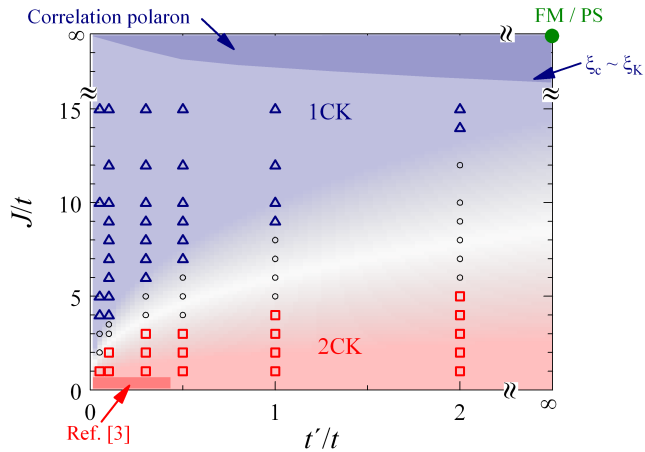}
\caption{
Ground-state phase diagram of the Kondo-polaron model, Eq.~\eqref{h}, obtained from DMRG
at $n_c=1$.
Triangles (squares) denote parameters with 1CK (2CK) behavior. The two phases are
separated by a QPT in the thermodynamic limit; in our finite-size numerics this
transition is smeared, with circles corresponding to parameters in the crossover region.
The perturbative arguments in favor of 2CK~\cite{lama_prl} apply in the limit of small
$J/t$.
At large $J$, a correlation polaron forms, its size $\xic$ being larger than the Kondo
screening length $\xik$. For $t=0$ and any finite $t'/J$, kinetic ferromagnetism (FM) is
realized at $n_c=1$, while phase separation (PS) occurs away
from half-filling (schematically shown).
}
\label{fig:pd}
\end{figure}

The purpose of this paper is a discussion of the full parameter space of the model
\eqref{h}, beyond the weak-coupling limit considered in Ref.~\onlinecite{lama_prl}. To
this end, we combine the analysis of various strong-coupling limits with comprehensive
numerical studies.
Our central result is that a QPT generically separates a small-$J$ phase with 2CK
screening of the impurity spin~\cite{lama_prl} from a 1CK phase at stronger coupling $J$, as
summarized in the phase diagram in Fig.~\ref{fig:pd}.
The transition in Eq.~\eqref{h} thus involves a change from local non-Fermi liquid to
Fermi-liquid behavior upon increasing $J$, accompanied by a jump in the residual impurity
entropy from $\ln \sqrt{2}$ to 0 \cite{aflu91}.
The transition is driven by varying only impurity parameters, $J$ or
$t'$, while keeping the bath parameters fixed, and corresponds to a hitherto unknown QPT.
%

In addition to the 1CK--2CK transition, we uncover an interesting strong-coupling regime,
where the motion of the impurity locally suppresses charge fluctuations in the electron gas,
thereby generating a ``correlation cloud'' (or ``correlation polaron'') whose size, $\xic$, is
dictated by kinetic energy and can be much larger than that of the Kondo screening cloud, $\xik$.

In the body of the paper, we present general arguments and numerical results from Density
Matrix Renormalization Group (DMRG) which lead to the above conclusions.
We also discuss possible realizations of the phenomena in the field of ultracold atomic
gases; we note that a related spin-only impurity problem has been recently studied using cold
bosonic atoms in a 1d optical lattice~\cite{kuhr13}.


{\it Weak-coupling limit, $\TKZ \ll t' \ll t$.}
We begin by summarizing the physics of the model \eqref{h} in the limit of small $J$,
discussed in Ref.~\onlinecite{lama_prl}. We use $\TKZ$ as a short-hand for the Kondo
temperature of a static impurity coupled with exchange $J$ to a band of
width $W$; for $J \ll t$ we have $\ln(\TKZ/W) \propto -W/J$.
The fate of the magnetic moment can be accessed in an expansion in $J$ around the
decoupled $J=0$ fixed point. This expansion is similar to the standard weak-coupling
expansion in the Kondo model, with the key difference that the recoil energy of the
impurity renders $2k_F$ backscattering processes from $J$ to be absent from the
low-energy sector. Hence, only processes with small momentum transfer -- involving either
left movers near $(-k_F)$ or right movers near $k_F$ -- contribute to the logarithmic
singularities, leading to flow equations equivalent to that of the 2CK effect. In other
words, the motion of the Kondo impurity causes left-moving and right-moving $c$ fermions
to form two separate screening channels for the impurity spin.

Importantly, this argument in favor of 2CK physics requires both $\TKZ \ll t'$, as
otherwise the recoil is too small to be relevant, and $\TKZ \ll t$, as otherwise the
separation into left movers and right movers is not justified.


{\it Strong-coupling limit, $J\gg t\gg t'$.}
In the limit $J\to\infty$ the impurity electron locks into a singlet with one conduction
electron. For $t'=0$ this singlet is immobile and effectively cuts the 1d electron gas.
From first-order perturbation theory in $t'$ one finds that
the singlet forms a Bloch wave with a kinetic energy of order $t'$.
Clearly, this corresponds to a slowly moving 1CK polaron of minimal size, i.e., 1CK
physics is realized in this limit.


{\it Strong-coupling limit, $J\gg t' \gtrsim t$.}
It is interesting to discuss the evolution of the singlet polaron upon increasing $t'/t$.
Whereas for $t'/t\to 0$ the conduction electrons simply adjust to the position of the
singlet, the case $t' \gtrsim t$ implies a faster motion of the polaron which is only
possible (without breaking the singlet) along a sequence of singly occupied $c$ sites. As
a result, the $c$-electron kinetic energy will be quenched in a vicinity of size $\xi_c$
of the impurity. Within this {\em correlation polaron} the impurity moves with a kinetic
energy of order $t'$, while the polaron itself -- consisting of the singlet surrounded by
singly occupied $c$ sites --  is a heavy object with kinetic energy of order $t$ (in a
manner similar to the ferromagnetic Kondo polaron described in
Ref.~\onlinecite{batista}). A variational estimate, assuming an immobile correlation
polaron, yields $\xic \propto t'/t$.
%
Thus, the correlation polaron emerges from the competition of impurity and $c$-electron
kinetic energies in the large-$J$ limit.
%


{\it Nagaoka limit, $t=0$.}
For completeness, we also mention the case of immobile $c$ electrons, $t=0$. Consistent
with the above discussion, $\xi_c\to\infty$ in this limit, i.e., the motion of the
impurity electron prefers singly occupied $c$ sites in the entire system. While the spin
alignment on the $c$ sites can be arbitrary for $J=\infty$, ferromagnetic alignment is
preferred for any finite $t'/J$ -- this kinetic magnetism can be understood as
double-exchange or Nagaoka ferromagnetism. If $n_c$ deviates from
half-filling, the system consequently phase-separates into a half-filled ferromagnetic
region and a region where $\langle n_i \rangle \neq 1$.

For both $t$ and $t'$ finite and small compared to $J$, the tendency towards
ferromagnetic alignment survives {\em inside} the correlation polaron. A detailed study
of this interesting regime is beyond the scope of this paper.


{\it Expected QPT.}
As argued above, 2CK screening is realized for $\TK \ll {\min}(t',t)$ where $\TK$ is now
a Kondo temperature in the presence of $t'$. On the other hand, the 1CK state of an
immobile impurity ($t'=0$) can be expected to be stable against small $t'\ll\TK$ (the
1CK polaron simply starts to move). Hence, a transition from 2CK to 1CK will occur upon
increasing $J$ or decreasing $t'$, as indeed confirmed by our numerics,
Fig.~\ref{fig:pd}.


{\it DMRG results.}
We have studied the model \eqref{h} using the DMRG technique~\cite{dmrg1,dmrg2} on finite
systems with $2\times L$ lattice sites. As the open boundary conditions commonly used with DMRG
lead to boundary pinning of the impurity electron, we have instead used anti-periodic boundary
conditions (APBC). This limits the maximum system size to $L=40$, which in turn implies
that Kondo screening with small $\TK<10^{-2}t$ will be hard to observe as the screening
cloud is much larger than the system size~\cite{kondodmrg}.
Unless otherwise noted, we have performed calculations varying $t'/t$, $J/t$, $h/t$ and
$L$; for details of the DMRG calculations see Ref.~\onlinecite{suppl}. In the
interest of numerical stability, most runs were done at $n_c=1$, but we have checked for
selected $J/t$ and $t'/t$ that our conclusions remain robust also for $n_c\neq 1$.

\begin{figure}[!t]
\includegraphics[height=0.95\columnwidth,angle=-90]
{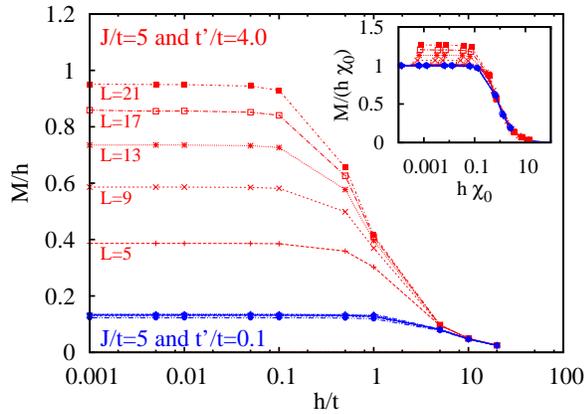}
\caption{
DMRG results for the impurity magnetization, plotted as $M/h$, as function of $h/t$ for
$L=5,9,13,17,21$.
Data are shown for $J/t=5$, $t'/t=0.1$ (blue) and $t'/t=4$ (red),
corresponding to the 1CK and 2CK phases, respectively.
The inset shows the same data as $M/(\chi_0 h)$ vs.\ $\chi_0 h$ where $\chi_0$ has been
obtained from a fit at large fields to Eq.~\eqref{mh_1ck}, for details see text. Lines
are guides to the eye.
} \label{fig:chih}
\end{figure}

\begin{figure}[!t]
\includegraphics[height=0.95\columnwidth,angle=-90]
{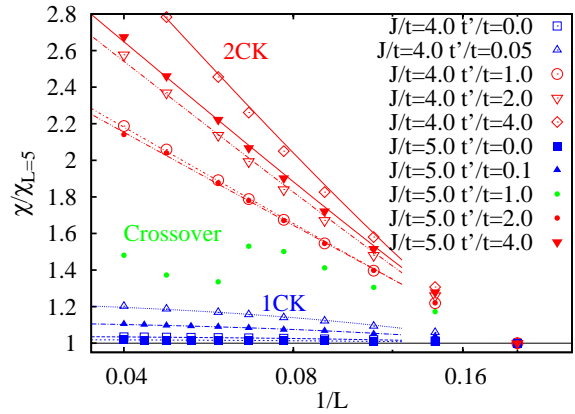}
\caption{
DMRG results for the local susceptibility $\chi$ (divided by its value at
length $L=5$) 
as function of inverse system size $1/L$ for parameter sets with $J/t=4$ and 5 and various $t'/t$.
While the data at small $t'$ show a clear saturation as $L\to\infty$, indicative of 1CK
(blue), $\chi$ at larger $t'$ increases logarithmically, consistent with 2CK (red). The
lines represent fits to Eqs.~\eqref{eq:fit1ck} and \eqref{eq:fit2ck}, for details see
text. One data set in the crossover region is also shown (green).
}
\label{fig:chil}
\end{figure}

The key quantity in our analysis is the impurity magnetization, $M=\sum_i \langle S_i^z
\rangle/L$, as function of applied impurity field, $h$, and system size $L$.
Sample data for $M/h$ is shown in Fig.~\ref{fig:chih}.
In all cases, $M \propto h$ as $h\to 0$ which allows us to define the
local impurity susceptibility, $\chi=M/h|_{h\to 0}$. This quantity is seen to strongly
depend on system size for small $J$.
Indeed, for the standard case of an immobile impurity, the finite-size behavior of the
susceptibility distinguishes 1CK and 2CK Kondo effects:
$\chi$ approaches a constant in the 1CK case, $\chi \propto 1/\TK$, whereas it
diverges logarithmically with system size in the
2CK case, $\chi \propto (1/\TK) \ln (\TK L)$.
The same qualitative behavior can be expected for mobile Kondo polarons -- this is well
borne out by our numerics: The data for $\chi$ in Fig.~\ref{fig:chil}
clearly show log-divergent $\chi(1/L)$ for large $t'$ and constant $\chi(1/L)$ for small
$t'$.

The distinct behavior at small and large $J$ is further illustrated in the inset of
Fig.~\ref{fig:chih}, where the data points at fixed $L$, $J/t$, $t'/t$ and high fields,
$h/t > 2$, are fitted to the 1CK strong coupling expression
\cite{kondodmrg},
\begin{equation}
M(h) = \frac{\chi_0 h}{\sqrt{1 + 4(\chi_0 h)^2}} \,,
\label{mh_1ck}
\end{equation}
and then plotted as $M/(\chi_0 h)$ vs.\ $\chi_0 h$. The large-$J$ data follow
Eq.~\eqref{mh_1ck}, again indicative of 1CK,
whereas the small-$J$ data systematically deviate at small $h$, with a deviation
increasing with increasing $L$.

To make the finite-size analysis of the susceptibility quantitative, we fit our DMRG data
for $\chi(L)$ utilizing the following crossover formulas~\cite{zarand00,suppl}:
\begin{equation}
\label{eq:fit1ck}
\chi_{\rm 1CK} = \frac{2\pi^2\Gamma+\Delta_L}{2(\pi^2\Gamma+\Delta_L)^2}
\end{equation}
and
\begin{equation}
\label{eq:fit2ck}
\chi_{\rm 2CK} = \frac{1}{2(\Delta_L+4\pi^2\Gamma)} \ln \left[1+ \frac{4\Gamma\Delta_L+(4\pi\Gamma)^2}{\Delta_L^2} \right].
\end{equation}
Here, $\Gamma$ is an energy scale proportional to the
(polaron) Kondo temperature, and $\Delta_L=b/L$ parameterizes finite-size effects on the
level spectrum.
The formulas have been adopted from a finite-size bosonization analysis of the Kondo
problem of an immobile impurity \cite{zarand00}; in this case $\Delta_L$ represents the
bath level spacing, with $b=4\pi t$ in the $L\to\infty$ limit.
Here we assume that Eqs.~\eqref{eq:fit1ck} and \eqref{eq:fit2ck} provide reasonable
descriptions of the data in the mobile-impurity case and for small $L$ as well, but we
treat $b$ as a second fit parameter, $b=b(t'/t,J/t)$, in addition to $\Gamma$.

Fitting $\chi(L)$ for all parameter sets (characterized by fixed values of $J/t$, $t'/t$, $n_c=1$)
to both Eqs.~\eqref{eq:fit1ck} and \eqref{eq:fit2ck} we observe the following:
(i) Some data sets can be fitted well by only one of the two forms, allowing us to
immediately discriminate between 1CK and 2CK behavior -- this mainly applies if the data
cover a range of $\Delta_L/\Gamma=0.1\ldots 1$.
(ii) Other data sets can be fitted by both forms, but often at the expense of extreme
values of the fitting parameters. In particular, $b/t\ll 1$ occurs when attempting to fit
large-$J$ data with the 2CK form Eq.~\eqref{eq:fit2ck}.
The evolution of the fitting parameters with $J$ and $t'$ is non-monotonic which allows
us to distinguish two regimes which clearly show 1CK and 2CK behavior, respectively~\cite{suppl}.
(iii) At intermediate values of $J$, we observe data sets which are not well fitted with
either of the two forms. Given that a putative QPT between 1CK and
2CK phases will be smeared for finite $L$, such behavior is consistent with the existence
of a quantum critical transition regime. This interpretation is supported by our
observation of significantly impaired convergence in this regime, which can be ascribed
to long-range entanglement which cannot be well captured by the matrix product states
underlying DMRG.

\begin{figure}[!t]
\includegraphics[height=0.95\columnwidth,angle=-90]
{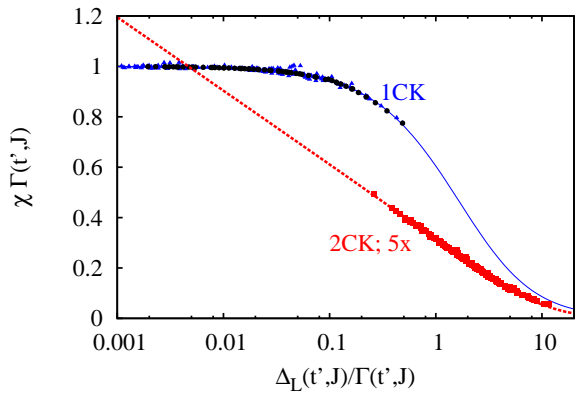}
\caption{
Scaling collapse of the DMRG results for the impurity susceptibility $\chi(L;J/t,t'/t)$
in the 1CK and 2CK phases. In each of the phases, $\chi \Gamma$ follows a universal
behavior as function of the finite-size parameter $\Delta_L/\Gamma$ where $\Delta_L=b/L$,
and $\Gamma$ and $b$ are fit parameters for each pair of $J/t$ and $t'/t$.
The symbols represent {\em all} data sets in Fig.~\ref{fig:pd} which could be uniquely
associated with either 1CK (blue) or 2CK (red) behavior; the black symbols corresponds to
$t'=0$.
The lines represent the scaling curves according to Eqs.~\eqref{eq:fit1ck} and
\eqref{eq:fit2ck}; the data deviate from these at small $L$.
}
\label{fig:scaling}
\end{figure}

The existence of two distinct screening regimes, together with the quality of the fits,
is demonstrated in Fig.~\ref{fig:scaling}, which shows universality of $\chi(L)$ when
plotted as $\chi\Gamma$ vs.\ $\Delta_L/\Gamma$. Here, we have shown those data sets which
could be uniquely assigned to either the 1CK or the 2CK phase; deviations from
universality occur for the data sets in the crossover region, for details see
Ref.~\onlinecite{suppl}.

\begin{figure}[!t]
\includegraphics[height=0.95\columnwidth,angle=-90]
{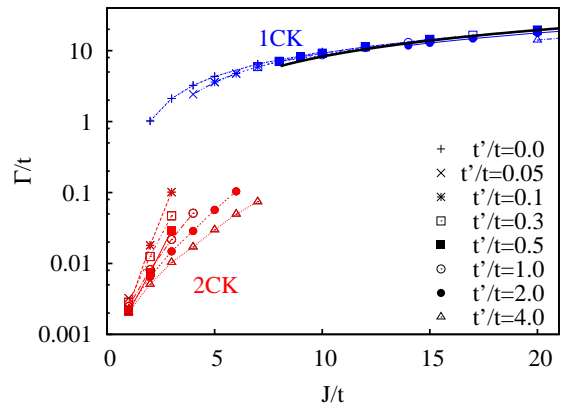}
\caption{
Kondo energy scale $\Gamma$, as extracted from the fits of $\chi$ to
Eqs.~\eqref{eq:fit1ck} and \eqref{eq:fit2ck}, as function of $J/t$ for different values
of $t'/t$ for the 1CK (2CK) regimes, shown in blue (red). The black line corresponds to
$\Gamma=J$ indicating the strong-coupling behavior. Data points with $\Gamma<10^{-2}t$
are subject to severe finite-size effects as $\xik\gg L$ here.
}
\label{fig:gamma}
\end{figure}

The fit parameter $\Gamma$, reflecting the polaron Kondo temperature $\TK$, is plotted in
Fig.~\ref{fig:gamma}.
First, we observe that 1CK behavior is seen for $\Gamma > \min(t,t')$ as anticipated,
whereas 2CK behavior is seen otherwise.
Second, $\Gamma$ becomes exponentially small for small $J$ and is proportional to $J$
for large $J$, as typical for the Kondo effect.
Third, $\Gamma$ is found to decrease with increasing $t'$ in the 2CK regime. This can be
rationalized by the fact that the motion of the Kondo polaron requires a spatial
adjustment of the screening cloud which tends to suppress screening. In contrast,
$\Gamma$ is weakly dependent on $t'$ in the 1CK regime, because here $\Gamma>t'$, i.e.,
the polaron moves sufficiently slowly for the screening cloud to adjust.

{\it Conclusions.}
We have established that a spinful particle moving in a 1d Fermi gas
displays a novel QPT between two phases with one-channel and
two-channel Kondo screening of the particle's spin.
While earlier perturbative arguments in favor of 2CK behavior apply to small Kondo
coupling $J$ only, $\TK \ll {\min}(t',t)$, our numerical results give evidence both for
1CK behavior at larger $J$ and for a transition to 2CK upon decreasing $J$.
Finding the universal field theory for this QPT is an interesting open issue.


The model in Eq.~\eqref{h} can in principle be realized using two species of atoms (with
two hyperfine states each) in an optical lattice~\cite{kondo_cold}. Due to the trapping
potential, left and right movers will be coupled, such that the transition between 2CK
and 1CK turns into a crossover. A suitable distinction between the two regimes is given
by the low-temperature mobility, which follows $T^{-2}$ ($T^{-4}$) in the 2CK (1CK)
case~\cite{lama_prl}.
Alternatively, dilute spinful holes in doped semiconductor nanowires \cite{calleja} can realize
the model Eq.~\eqref{h}. Here, the change from local non-Fermi liquid to Fermi liquid
behavior may be detected using the magnetic response in a Zeeman field,
which is singular (regular) in the 2CK (1CK) case \cite{zarand00,coxzawa,ek}.

Future work could possibly investigate the non-equilibrium dynamics near the Kondo-polaron QPT as well
as the influence of bath interactions, i.e., the physics of Kondo polarons in a Luttinger
liquid.


We thank B. Alascio, P. Cornaglia, A. Feiguin, E. Fradkin, T. Giamarchi, and C. Vojta for discussions.
This research was supported by the DFG (FG 960) and the German-Israeli
Foundation.
M.V.\ also acknowledges support by the Heinrich-Hertz-Stiftung NRW and the
hospitality of the Centro At\'omico Bariloche where part of this work was performed.


\end{document}


\title{
Supplementary information for:\\
Quantum phase transition between one-channel and two-channel Kondo polarons 
}

\author{Juli\'an Rinc\'on}
\affiliation{Center for Nanophase Materials Sciences, Oak Ridge National Laboratory, Oak Ridge, TN 37831, USA}
\author{Daniel J. Garc\'{\i}a}
\author{K. Hallberg}
\affiliation{Instituto Balseiro, Centro At\'omico Bariloche, CNEA and CONICET, 8400 Bariloche, Argentina}
\author{Matthias Vojta}
\affiliation{Institut f\"ur Theoretische Physik, Technische Universit\"at Dresden, 01062 Dresden, Germany}

\date{\today}
\maketitle


\section{Derivation of finite-size fitting formulae}
\label{sec:deriv}

In this section, we shall derive the finite-size scaling formulae for both the
one-channel (1CK) and the two-channel (2CK) Kondo models, which are used in the paper for
the analysis of the susceptibility. The formulae are based on the calculation of the
finite-size crossover spectrum in a model with linearized bath dispersion using
bosonization and refermionization, as presented by Zarand and von Delft.\cite{zarand00}
For simplicity, all considerations will be restricted to the particle--hole symmetric
case, which corresponds to $n_c = 1$ in the model Eq.~\eqh\ of the main text. Due to the
assumed linear dispersion, the results are strictly valid for a lattice model only in the
limit $\TK \ll t$.

\subsection{Two-channel Kondo formula}

The local susceptibility $\chi$ can be defined as the second derivative w.r.t. the
impurity field $h$ of the ground state energy of a diagonalized refermionized Hamiltonian
form of the original 2CK model.
According to Eqs.~(53), (61), and (62) in Ref.~\onlinecite{zarand00}, $\chi$ can be expressed
as
\begin{equation}
\chi = -\frac{1}{2}\sum_{\varepsilon\ge 0} \frac{\partial^2\varepsilon}{\partial h^2},
\end{equation}
where $\varepsilon$ are excitation energies, which can be obtained as the positive roots of the
transcendental equation
\begin{equation}
\label{vareps}
\frac{\varepsilon 4\pi\Gamma}{\varepsilon^2-\varepsilon_d^2} = -\cot \pi
(\varepsilon/\Delta_L - P/2).
\end{equation}
Here, $\varepsilon_d = h$ and $\Gamma$ are the level position and width of the effective
resonant level model obtained after refermionization.\cite{ek} $\Delta_L=2\pi/L$ is a
level spacing of the bath with Fermi velocity set to unity, and $P=0$ or 1 if the total
number of electrons is even or odd.

Taking an implicit derivative, we find that $\partial_h \varepsilon|_{h=0}\equiv 0$.
Differentiating twice, setting $h=0$, and using standard trigonometric identities, we
find
\begin{equation}
\label{chicont}
\chi = -\int_{0}^\infty \frac{d\varepsilon}{\varepsilon} \frac{4\pi\Gamma}{4\pi\Gamma\Delta_L + \pi \left[(4\pi\Gamma)^2 + \varepsilon^2\right] }.
\end{equation}
Here we have switched to the continuum limit by replacing the sum by an integral,
$\sum_{\varepsilon\ge 0}\to (1/\Delta_L)\int d\varepsilon$, assuming that the levels
$\varepsilon$ are approximately spaced by $\Delta_L$, see Eq.~(65) of
Ref.~\onlinecite{zarand00}.
The expression \eqref{chicont} is, in a sense, the
continuous version of Eq.~(75) in Ref.~\onlinecite{zarand00}. Notice that this result is
valid for both periodic and open boundary conditions as established by
Eq.~\eqref{vareps}.

The integral in Eq.~\eqref{chicont} diverges logarithmically for small arguments. A natural
cutoff is given by the level spacing $\Delta_L$ itself.
Performing the integral with lower limit $\Delta_L$ yields
\begin{equation}
\chi_{\rm 2CK} = \frac{1}{2(\Delta_L+4\pi^2\Gamma)} \ln \left[1+ \frac{4\Gamma}{\Delta_L}+\left(\frac{4\pi\Gamma}{\Delta_L}\right)^2 \right].
\label{fit2ck}
\end{equation}
This expression gives the desired finite-size behavior of $\chi$ at zero temperature and
is identical to Eq.~\eqfittc\ of the main text. Note that Eq.~\eqref{fit2ck} yields the
exact asymptotic behavior of $\chi$ in the limit large-system limit, $\Gamma \gg
\Delta_L$, see Eq.~(76) of Ref.~\onlinecite{zarand00}; for $\Gamma \lesssim \Delta_L$ it
is an approximation because of our rough treatment of the lowest excitation energies
$\varepsilon$.

\subsection{One-channel Kondo formula}

As for the 2CK case, we can also calculate the finite-size behavior of $\chi$ in the case
of the 1CK model. Based on a similar expression to the 2CK case for the excitation energy
$\varepsilon$, derived in appendix D of Ref.~\onlinecite{zarand00}, one obtains:
\begin{equation}
\chi = -\sum_{\varepsilon< 0} \frac{\partial^2\varepsilon}{\partial h^2},
\end{equation}
where the excitation energies $\varepsilon$ are now the negative roots of the transcendental equation
\begin{equation}
\frac{\pi\Gamma}{\varepsilon-h}=-\cot \pi\varepsilon/\Delta_L\,,
\end{equation}
see Eq.~(D37) of Ref.~\onlinecite{zarand00}.
As in the previous subsection, differentiating twice with respect to $\varepsilon$, using
trigonometric identities, and setting $h=0$, we find that
\begin{equation}
\chi = -\int_{-\infty}^0 d\varepsilon\, \frac{2\Gamma\varepsilon\left[(\pi\Gamma)^2 + \varepsilon^2\right]}{\left( \Gamma\Delta_L + \left[ (\pi\Gamma)^2 + \varepsilon^2 \right] \right)^3}.
\end{equation}
Notice that in this case the integral is convergent and no regularization is required.
Performing the integral we find
\begin{equation}
\chi_{\rm 1CK} = \frac{2\pi^2\Gamma+\Delta_L}{2(\pi^2\Gamma+\Delta_L)^2},
\label{fit1ck}
\end{equation}
which describes the finite-size scaling of $\chi$ and is identical to Eq.~\eqfitoc\ of
the main text.
In the large-system limit, $\Gamma\gg\Delta_L$, we find
\begin{equation}
\chi_{\rm 1CK} \simeq \frac{1}{\pi^2\Gamma} - \frac{3}{2}\frac{\Delta_L}{(\pi^2\Gamma)^2} + \mathcal{O}(\Delta_L^2).
\end{equation}
The leading term recovers the standard
expression $\chi\propto 1/\TKZ$ since $\Gamma\sim \TKZ$, the Kondo temperature.

\section{DMRG implementation}

In the following, we shall discuss the main technical aspects of our DMRG results on the
dispersive magnetic impurity.

The DMRG results were obtained for $2\times L$ chains, with odd $L$ from $5$ up to $37$
sites, with antiperiodic boundary conditions and a half-filled $c$-electron band such that
the ground state of Eq.~\eqh\ in the main text is always a singlet, and allows us to work
in the subspace $S^z_{\rm tot} = 0$. The largest truncation error in the calculations was
$\epsilon_\rho\sim 10^{-6}$, a remarkable value considering the closed boundary
conditions. Such an error translates to keeping between $512$ and $2048$ states per block
and performing $12$ to $18$ sweeps in the finite-size algorithm.

\subsection{Benchmark}

In order to benchmark our DMRG algorithm, we have reproduced Wang's
results\cite{kondodmrg} for a static Kondo impurity coupled to an interacting
one-dimensional system.

For a large variety of parameter values of the mobile impurity Hamiltonian, Eq.~\eqh\ of the
main text, we have also cross-checked our results against exact diagonalization calculations
finding excellent agreement for lattices up to $2\times 11$ sites.

\subsection{Magnetization and susceptibility}

The magnetic field is applied only to the impurity and the magnetization, $M$, is obtained as the mean value of the $z$ component of the impurity spin operator, 
\begin{equation}
M = \frac{1}{L}\sum_{i=1}^L\langle\psi_0(h)| S^z_i |\psi_0(h)\rangle,
\end{equation}
where $|\psi_0(h)\rangle$ is the ground state of Eq.~(1) of the main text obtained with
DMRG, at a given field $h$ and system size of $2\times L$ sites. The susceptibility
$\chi$ is approximated as the ratio of magnetization and applied field at a finite, albeit small,
field, $\chi := M/h$ at $h/t=10^{-3}$; we have checked that smaller values of $h$ deliver
essentially identical results for the system sizes considered here.

\begin{figure}[!t]
\includegraphics*[height=0.85\columnwidth, angle=-90]{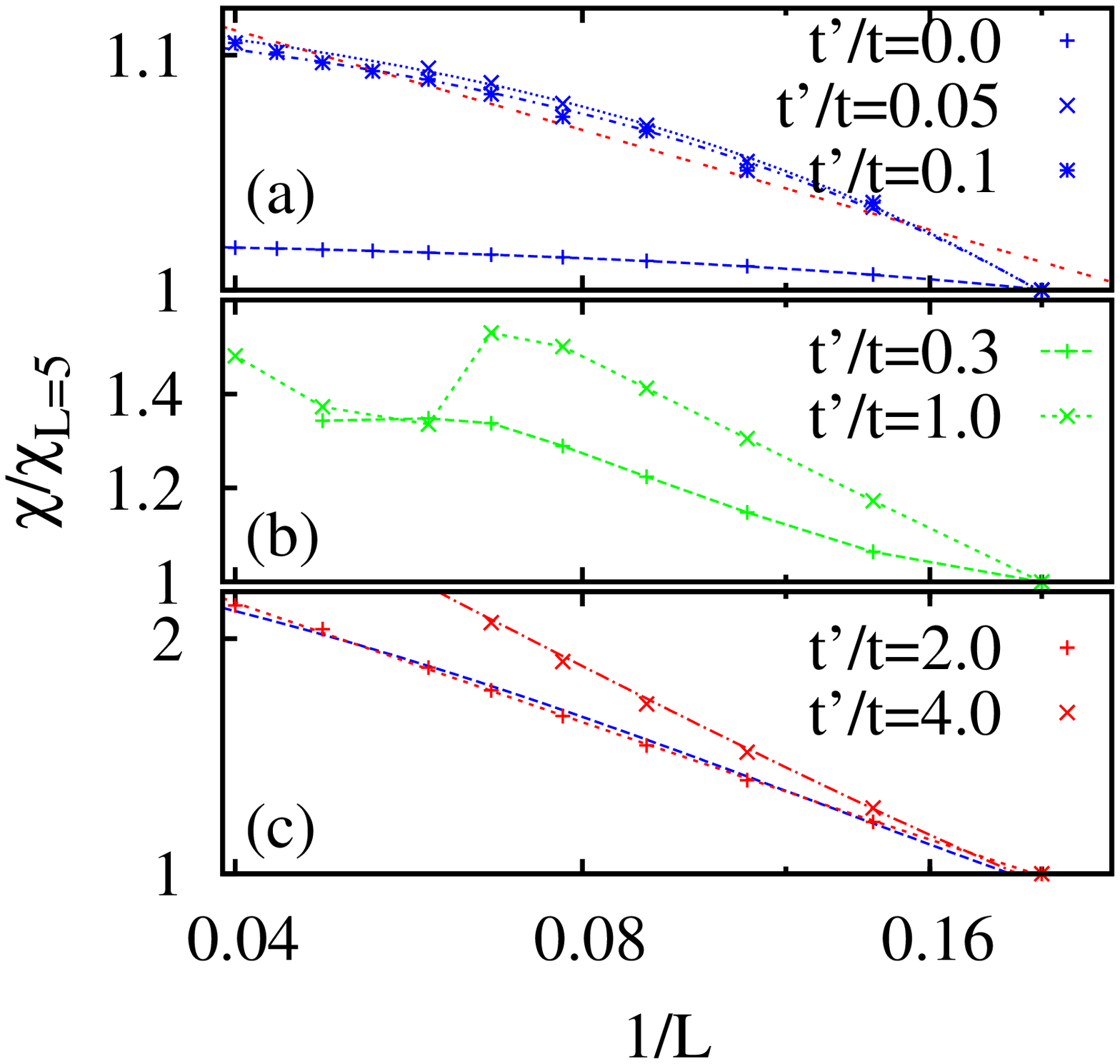}
\caption{
DMRG results for the local susceptibility $\chi$ normalized to its value at length $L=5$
as a function of $1/L$. Shown are data sets with $J/t=5$ and varying $t'/t$ which belong
to the (a) 1CK, (b) crossover, and (c) 2CK regimes. Fits using Eq.~\eqref{fit2ck}
(Eq.~\eqref{fit1ck}) are shown as a red (blue) lines; dashed lines in (c) are guides to
the eye. In panel (a) the 2CK fit is for $t'/t=0.1$, in panel (c) the 1CK fit is for
$t'/t=2.0$.
}
\label{fig:Srawdata}
\end{figure}

\section{Additional results}

We now present selected additional results for the local susceptibility of the Kondo
polaron model, obtained by DMRG, which illustrate the crossover between 1CK and 2CK
behavior.

\subsection{Crossover at fixed $J/t$}

Figure~\ref{fig:Srawdata} shows data of the susceptibility $\chi$ scaled to its value at
length $L=5$ as a function of the inverse system size for $J/t=5$. Upon increasing $t'$,
the data can be clearly separated in three different regimes, namely 1CK in
Fig.~\ref{fig:Srawdata}(a), crossover in Fig.~\ref{fig:Srawdata}(b), and 2CK in
Fig.~\ref{fig:Srawdata}(c).

\begin{figure}[!t]
\includegraphics*[width=\columnwidth]{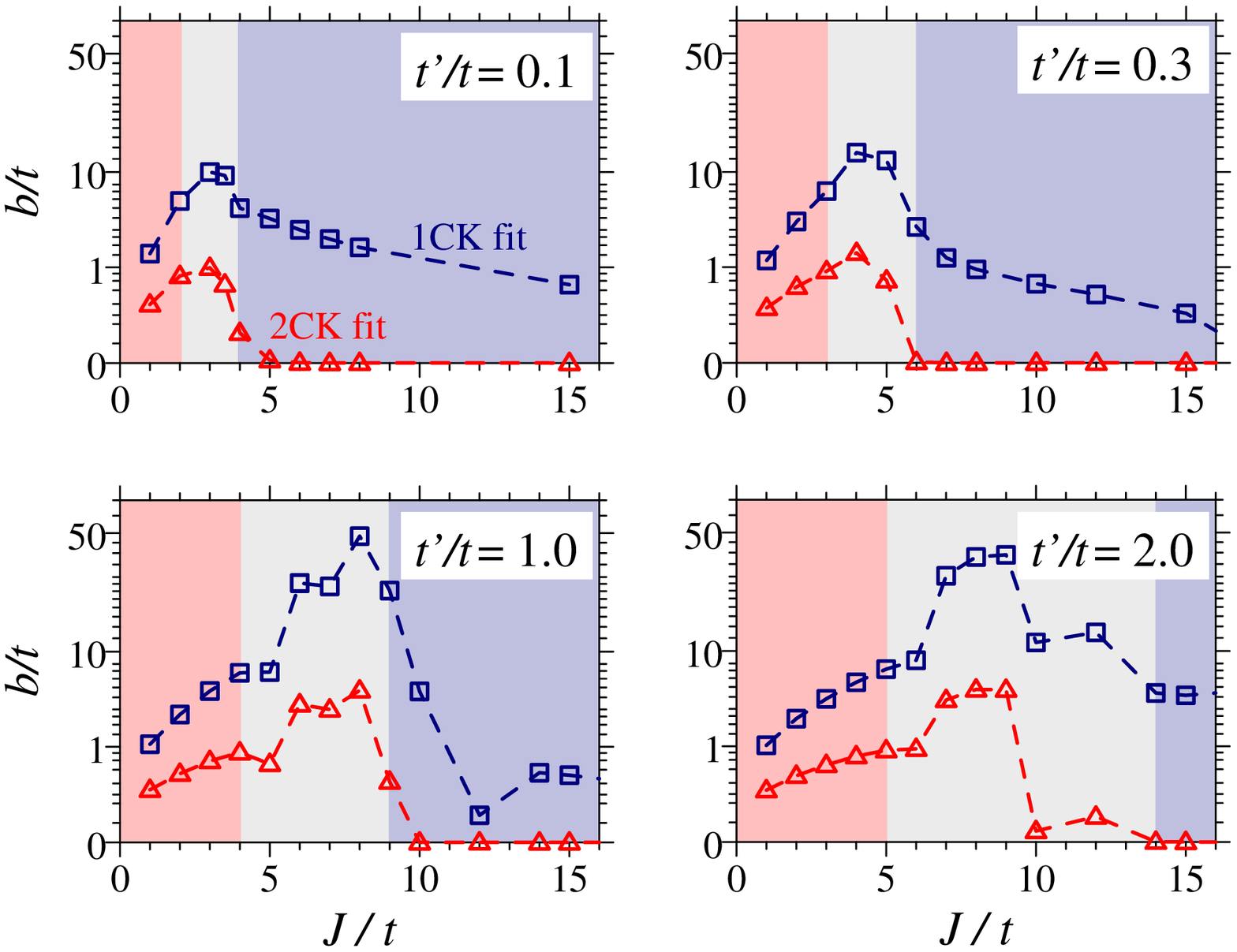}
\caption{
Evolution of the fitting parameter $b$ with $J/t$ from the finite-size fitting
formulae Eqs.~\eqref{fit2ck} and ~\eqref{fit1ck} for the 1CK and 2CK cases, for several
values of $t'/t$.
The maximum in $b(J)$ separates the regions of 1CK (blue shaded) and 2CK (red shaded)
behavior, with an extended crossover region in between (gray shaded). The regimes
are identical to the ones shown in the phase diagram in Fig.~\figpd\ of the main paper.
}
\label{fig:Sfigb}
\end{figure}

In panel (a), the 1CK fits using Eq.~\eqref{fit1ck} are excellent, while the 2CK fits
(one is shown) are inferior. In addition, the 2CK fits yield extremely small values of
the fit parameter $b$, see also Fig.~\ref{fig:Sfigb}, which would correspond to an
unphysically small bath level spacing and thus indicate the failure of the fit.

Similarly, in panel (c), the 2CK fits via Eq.~\eqref{fit2ck} are excellent and reproduce
the logarithmic divergence of the susceptibility. In contrast, the 1CK fits, although
reasonably accurate for the $L$ range available, display a downward curvature not present
in the data.

In the crossover regime, neither Eq.~\eqref{fit2ck} nor Eq.~\eqref{fit1ck} are suitable
fits -- in the range of system sizes studied -- to any of the results shown in
Fig.~\ref{fig:Srawdata}(b).

\subsection{Evolution of fit parameters}

As described in the main text, we have fitted the $\chi(L)$ data for all data sets
characterized by fixed $J/t$, $t'/t$, and $n_c=1$ to {\em both} finite-size formulae
\eqref{fit2ck} and \eqref{fit1ck}. This yields $b$ and $\Gamma$ as function of $J/t$ and
$t'/t$, separately for the 1CK and 2CK cases.

In Figure~\ref{fig:Sfigb}, we show the $J$ evolution of the resulting fit parameter $b$,
defined via the level-spacing parameter $\Delta_L=b/L$ in Eqs.~\eqref{fit2ck} and
\eqref{fit1ck}, for selected values of $t'/t$. As stated in the main text, $b=4\pi t$ is
expected for an immobile impurity from the derivation in Sec.~\ref{sec:deriv}, but
deviations can arise due to (i) departures from the idealized linear bath dispersion
(i.e. if $\TK\ll t$ is not fulfilled), and (ii) recoil effects due to impurity motion.

As a check of our means of analysis, we have first considered the immobile-impurity case
$t'=0$ where 1CK behavior is known to occur in the present model. We find (data not
shown) that $b$ of the 1CK fit is of order unity and monotonically decreasing with
increasing $J$. Extrapolating to $J\to 0$, where $\TK \ll t$, we find $b/t\approx 10$
which is close to the expected value $4\pi$; note that the data are unreliable for
$J/t\lesssim 1$ where $\TK$ becomes smaller than $10^{-2}t$. In contrast, the $b$ extracted
from a 2CK fit takes unphysically small values, i.e., smaller than $10^{-2}t$ for $J/t\geq 2$,
indicating that a 2CK description is not applicable.

\begin{figure}[b]
\includegraphics*[height=\columnwidth,angle=-90]{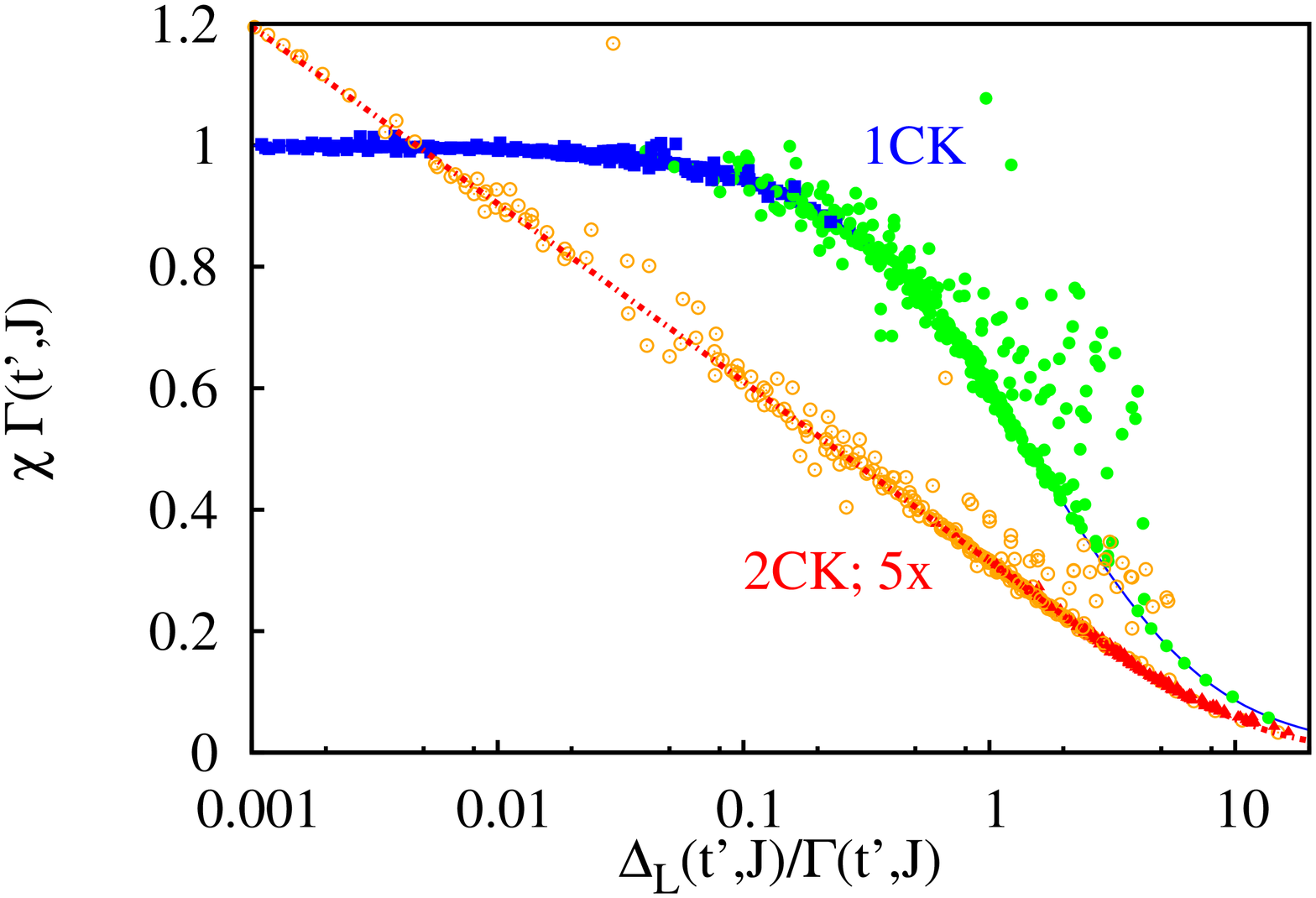}
\caption{
Scaling plot of the DMRG results for the susceptibility $\chi(L;J/t,t'/t)$ as in Fig.~4
of the main paper. In addition to the 1CK (blue) and 2CK (red) data sets, which display
universal behavior of $\chi \Gamma$ as function of $\Delta_L/\Gamma$, the data sets in
the crossover region are shown as well, which yield the green (orange) data points when
fitted to the 1CK (2CK) crossover formula.
The lines represent the scaling curves according to Eqs.~\eqref{fit2ck} and \eqref{fit1ck}.
}
\label{fig:Sscaling}
\end{figure}

For finite $t'$, Fig.~\ref{fig:Sfigb}, we observe a {\em non-monotonic} behavior of
$b(J)$, with a maximum at intermediate $J$. For larger $J$, e.g. $J/t\geq 4$ in
Fig.~\ref{fig:Sfigb}(a), the 1CK values of $b$ are of order unity while the 2CK $b$
values become tiny. This is similar to the $t'=0$ case and allows us to assign this
regime with 1CK behavior, as is also expected from the strong-coupling arguments
presented in the main text.
In contrast, for small $J$, e.g. $J/t\leq 2$ in Fig.~\ref{fig:Sfigb}(a), $b$ is of order
unity for both the 1CK and 2CK fits. The superior quality of the 2CK fits,
Fig.~\ref{fig:Srawdata}(c), together with the weak-coupling arguments of
Ref.~\onlinecite{lama_prl}, point to an identification of this as the 2CK regime.

Interestingly, the crossover region, as identified by the low quality of both 1CK and 2CK
fits, Fig.~\ref{fig:Srawdata}(b), precisely corresponds to the region around the maximum
of $b(J)$ for fixed $t'/t$, Fig.~\ref{fig:Sfigb}. Taken together, this underlines that this
crossover region separates regions of well-defined 2CK and 1CK behavior, as described in
the main text.

\subsection{Deviations from scaling}

Fig.~4 of the main text reports an approximate scaling collapse of the susceptibility
$\chi(L;J/t,t'/t)$, separately for the 1CK and 2CK phases, when plotted as $\chi \Gamma$
vs. $\Delta_L/\Gamma$ where $\Delta_L=b/L$.
Figure~\ref{fig:Sscaling} shows the same plot, but now with a complete set of scaled data
points in the following sense: In addition to those data sets which were unique assigned
to either the 1CK (blue squares) or 2CK (red triangles) regimes, we also show the
remaining data sets. These data sets, belonging to the crossover region, can be fitted
(with low quality) to both 1CK and 2CK crossover formulae, with the results shown as
filled green and open orange circles, respectively.

The 1CK-fitted crossover data (green) show clear departures from the expected universal
crossover curve. Such deviations are also present, albeit smaller, for the 2CK-fitted
crossover data (orange). The fact that the deviations are smaller reflects the fact that
the finite-size behavior of $\chi$ in the 2CK case is not very different from that of a
1CK case with $\xik>L$, see also Fig.~\ref{fig:Srawdata}(c).

\subsection{Quantum criticality}

According to the general quantum critical phenomenology,\cite{ssbook} the crossover
region in Fig.~\figpd\ of the main paper should be associated with a unique {\em quantum
critical} finite-size scaling, distinct from that of the 1CK and 2CK phases (unless the
QPT is of first order, for which we have no indications).

Unfortunately, the convergence problems which we encountered when applying DMRG in the
crossover region prevent us from acquiring accurate data near the putative QPT. The
present data, see e.g. Fig.~\ref{fig:Srawdata}(b), do not allow us to extract the quantum
critical scaling; therefore this is left for future work.
